\documentclass[aps,prd,superscriptaddress,12pt,showpacs,notitlepage]{revtex4}
\usepackage{graphics}
\usepackage{graphicx}
\usepackage{amsfonts}
\usepackage{amsmath}

\newcommand{\be}{\begin{equation}}
\newcommand{\ee}{\end{equation}}
\newcommand{\bea}{\begin{eqnarray}}
\newcommand{\eea}{\end{eqnarray}}
\newcommand{\pa}{\partial}
\newcommand{\bb}{\bibitem}
 
\begin{document}
\title{Twinlike models with identical linear fluctuation spectra}
\author{C. Adam}
\affiliation{Departamento de F\'isica de Part\'iculas, Universidad de Santiago de Compostela and Instituto Galego de F\'isica de Altas Enerxias (IGFAE) E-15782 Santiago de Compostela, Spain}
\author{J.M. Queiruga}
\affiliation{Departamento de F\'isica de Part\'iculas, Universidad de Santiago de Compostela and Instituto Galego de F\'isica de Altas Enerxias (IGFAE) E-15782 Santiago de Compostela, Spain}

\pacs{11.30.Pb, 11.27.+d}

\begin{abstract}
Recently, the possibility of so-called twinlike field theories has been demonstrated, that is, of different field theories which share the same topological defect solution with the same energy density. Further, purely algebraic conditions have been derived which the corresponding Lagrangians have to obey in order that the field theories be twins of each other. A further diagnostical tool which, in general, allows to distinguish the topological defects of a given theory from the corresponding defects of its twins is the spectrum of linear fluctuations about these defects. Very recently, however, explicit examples of twin theories have been constructed such that not only their shapes and energy densities coincide, but also their linear fluctuation spectra are the same. Here we show that, again, there exist purely algebraic conditions for the Lagrangian densities which imply that the corresponding field theories are twins and that the fluctuation spectra about their defects coincide. These algebraic conditions allow to construct an infinite number of twins with coinciding fluctuation spectra for a given theory, and we provide some explicit examples. The importance of this result is related to the fact that coinciding defects with coinciding energy densities and identical fluctuation spectra are almost indistinguishable physically, that is, indistinguishable in a linear or semiclassical approximation. 
This implies that the measurable physical properties of a kink, in general, do not allow to determine the theory which provides the kink uniquely. Instead, in principle an infinite number of possible theories has to be considered.

\end{abstract}

\maketitle 

\section{Introduction}
One of the most fertile concepts in theoretical physics in the last decades has been the concept of topological defects or topological solitons (see e.g. \cite{man-sut-book}). They are ubiquitous in condensed matter systems and, besides this, are deemed relevant for the cosmology of the early universe. Topological defects may, for instance, contribute to the structure formation in the very early universe (e.g., during or at the end of inflation) \cite{Vil}-\cite{Battye}. 
A topological soliton is, in general, a static solution of the Euler--Lagrange equations of the given field theory with finite energy which obeys nontrivial boundary conditions. Further, the stability of the topological soliton against transitions to the vacuum is guaranteed by the fact that a deformation to the vacuum configuration with trivial boundary conditions would require to change the field in an infinite volume and, therefore, cost an infinite amount of energy.  
The relevant data characterizing the physical properties of a soliton are, first of all, its shape or profile (i.e., the soliton solution itself), and its energy density. Additional important information is contained in the so-called spectrum of linear fluctuations about the topological defect. In order to determine this spectrum, one calculates the fluctuations about the soliton up to second order in the action (or up to first order in the Euler--Lagrange equations). For the fluctuation field then one introduces a temporal Fourier decomposition, which results in a stationary second order equation of the Schr\"odinger type. The (in general, infinitely many) solutions of this equation together with the allowed frequencies constitute the spectrum of linear fluctuations. The first relevant information contained in the spectrum of linear fluctuations is linear stability. For a stable soliton, the spectrum should contain no negative mode (i.e., no imaginary frequency). Another aspect where the fluctuation spectrum is important is the issue of semiclassical quantization in the presence of solitons \cite{Go-Jac} (for an easy to follow discussion see \cite{raja-book}). Concretely, the discrete part of the fluctuation spectrum describes some excited states of the soliton or, equivalently, soliton-meson bound states. Here by "meson" we mean a fluctuation field which is Gaussian in the leading approximation and obeys the boundary conditions of the vacuum configuration. Further, the continuous part of the spectrum describes soliton-meson scattering.   

The discussion so far has been for general soliton models, but now we want to restrict to the case of a real scalar field in 1+1 dimensions.
The standard scalar field theory in 1+1 dimensions is
\be \label{standard-lag}
{\cal L}_s = X - U(\phi) \quad , \quad X\equiv \frac{1}{2}\pa_\mu \phi \pa^\mu \phi
\ee
and we shall require that $U$ is nonnegative,
\be 
U(\phi)\ge 0 \quad \forall \quad \phi .
\ee
This theory may support topological solitons (kinks) provided that the potential $U$ has at least two vacua, i.e., there exist at least two (constant) values $\phi = \phi_i$ such that $U(\phi_i)=0$. A kink is a static solution $\phi_k (x)$ which, in general, interpolates between two adjacent vacua, i.e., $\phi_k (-\infty)=\phi_i$, $\phi_k (\infty)=\phi_{i+1}$.
The corresponding static kink equation is ($\phi ' \equiv \pa_x \phi$)
\be \label{kink-eq}
\frac{1}{2}\phi'^2 \equiv -X =U
\ee
with the two roots (for kink and antikink)
\be
\phi' = \pm \sqrt{2U} .
\ee
The kink equation (\ref{kink-eq}) results from the static second order Euler--Lagrange equation by performing one integration, where the integration constant must be set equal to zero in order to satisfy the kink boundary conditions. Finally, the linear fluctuation equation in the kink background may be derived by inserting the decomposition $\phi(t,x) = \phi_k (x) + \eta (t,x)$ and the temporal Fourier decomposition $\eta (t,x)= \cos (\omega t)\eta (x)$ into the Euler--Lagrange equation and keeping terms linear in $\eta$. Explicitly, the linear fluctuation equation reads
($U_{,\phi} \equiv \pa_\phi U$, etc.)
\be
-\eta '' = (\omega^2 - U_{,\phi\phi}\vert_{\phi_k} )\eta 
\ee
 where the notation $\vert_{\phi_k}$ means that the expression has to be evaluated for the kink solution. The solutions of this Schr\"odinger type equation together with the allowed frequencies $\omega$ determine the spectrum of linear fluctuations in this case. 

Up to now the logical line of reasoning has been to begin with a field theory and to derive from this starting point the topological defect (kink) and its properties. Now we want to see whether and how far this logical arrow can be reversed. That is to say, we start with a kink solution together with its properties, like energy density and linear fluctuation spectrum, and we want to know whether or to which degree we may recover the theory which gives rise to this defect solution with its properties. The answer depends on the class of Lagrangians we are willing to admit. For a standard scalar field theory (\ref{standard-lag}), the kink solution itself is already sufficient to recover the Lagrangian, i.e., the potential, by inverting the solution $\phi = \phi_k (x) \Rightarrow x= x_k(\phi )$ and by inserting the resulting expression into the kink equation,
\be
\phi'^2 (x)= \phi'^2 (x_k (\phi)) \equiv 2U(\phi) ,
\ee
which determines $U(\phi)$. On the other hand, the situation will be different if we allow for a more general class of Lagrangians. Concretely, we want to admit Lagrangians which are general functions of both $\phi$ and $X\equiv (1/2) \pa_\mu \phi \pa^\mu \phi$. There are several reasons which make these theories with a generalized kinetic term (the so-called K field theories)
worth considering. First of all, K field theories have been applied already to some problems in cosmology, like inflation (so-called K-inflation
\cite{k-infl}), late time acceleration (so-called K-essence \cite{k-ess}), or in the brane world scenario \cite{comp-brane} - \cite{brane-bazeia}. Secondly, generalized kinetic terms may serve to stabilize static field configurations, evading thereby the Derrick theorem and allowing the existence of soliton solutions. The third and probably strongest case in favor of K field theories is related to the fact that in many circumstances scalar field theories are interpreted as effective field theories which result from  the integration of UV degrees of freedom of some more fundamental theory. In this case of an effective field theory, higher powers of derivatives are induced naturally, and therefore they have to be taken into account. In this paper we are specifically interested in K field theories whose topological defects coincide with the standard ones, but let us mention, nevertheless, that K field theories in general give rise to a much richer phenomenology of possible topological defects \cite{babichev1}, \cite{nicole-mod}, like, e.g. solitons with compact support (so-called compactons) \cite{werle}  - \cite{fring}. Other more mathematical aspects of K field theories have been discussed, e.g., in \cite{bab-muk-1} and in \cite{bergli1}.

For the generalized dynamics of K field theories (i.e., for general Lagrangians ${\cal L}(X,\phi)$) it was found recently \cite{trodden} that different field theories may exist which share the same topological defect with the same energy density. The coinciding kinks with their coinciding energy densities were dubbed twin or Doppelg\"anger defects in \cite{trodden}, and the models which give rise to these identical kink solutions are called twinlike models. The investigation of twinlike models was carried further in \cite{twins-bazeia1} and in \cite{twins-1}. Specifically, in \cite{twins-1} it was demonstrated that there exist purely algebraic necessary and sufficient conditions for a Lagrangian ${\cal L}(X,\phi)$ to be the twin of a standard theory ${\cal L}_s = X-U$. As these conditions are algebraic, they do not require the knowledge of the topological defect solution and, therefore, allow the simple construction of an infinite number of twins for any given standard field theory supporting topological defects. 
Very recently, in \cite{twins-bazeia2} explicit examples of K field theories were found which not only are twin models of standard field theories, but where also the fluctuation spectra of the standard defect and its K field twins coincide, making the standard defect and its twins almost completely indistingushable physically. 
This implies that the measurable physical properties of a kink, in general, do not allow to determine the theory which provides the kink uniquely. Instead, in principle an infinite number of possible theories has to be considered.

It is the purpose of the present paper to show that, again,    
there exist purely algebraic conditions for a Lagrangian density which imply that the corresponding field theory is the twin of a standard scalar field theory {\em and} that the fluctuation spectra about their defects coincide. Further, these algebraic conditions allow to explicitly construct an infinite number of twins with coinciding fluctuation spectra for any given standard field theory. Concretely, in Sec. II we briefly review some known facts about twinlike models which we need. In Sec. III, we derive the algebraic conditions for coinciding fluctuation spectra and provide some explicit examples. Further we discuss the relation of our results with the examples of Ref. \cite{twins-bazeia2}. Finally, Sec. IV contains our conclusions.

\section{Twinlike models}
The algebraic twin conditions require the first order form of the static field equations, so let us briefly review this issue (for more details see, e.g., \cite{twins-1}, \cite{bazeia3}). 
For a general Lagrangian ${\cal L}(X,\phi)$ where $X\equiv \frac{1}{2} \pa_\mu \phi \pa^\mu \phi = \frac{1}{2}(\dot{\phi}^2 - \phi '^2)$, 
the Euler--Lagrange equation reads
\be \label{eul-lag}
\pa_\mu ( {\cal L}_{,X} \pa^\mu \phi ) - {\cal L}_{,\phi} =0 .
\ee
Further, the energy momentum tensor is
\be
T_{\mu\nu} = {\cal L}_{,X} \pa_\mu \phi \pa_\nu \phi - g_{\mu\nu} {\cal L}
\ee
which, for static configurations $\phi = \phi (x)$, $\phi ' \equiv \pa_x \phi$, simplifies to
\bea \label{stat-en-de}
T_{00} &=& {\cal E} = -{\cal L} \\
T_{11}  &=& {\cal P} = {\cal L}_{,X} \phi'^2 + {\cal L}
\eea
where ${\cal E}$ is the energy density and ${\cal P}$ is the pressure. The static Euler--Lagrange equation may be integrated once to give
\be \label{first-order-eq}
-2X{\cal L}_{,X} + {\cal L} \equiv {\cal P} = 0.
\ee
The general first integral allows for a nonzero constant on the r.h.s. (nonzero pressure), but the boundary conditions for finite energy field configurations require this constant to be zero (zero pressure condition).
For a standard field theory ${\cal L}_s= X-U$, the energy density and pressure read
\bea \label{stat-en-de-s}
 {\cal E}_s &=& -X+U = \frac{1}{2}\phi'^2 +U \\
 {\cal P}_s &=& -X-U = \frac{1}{2}\phi'^2 -U ,
\eea
and for a kink solution $\phi_k$ obeying $\phi'^2_k =2U$ these simplify to 
\bea \label{stat-en-de-s2}
 {\cal E}_s \vert_{\phi_k} &=&   -2X\vert_{\phi_k}  =2U\vert_{\phi_k} \\
 -{\cal P}_s &=& X+U \equiv 0 .
\eea
Obviously, a K field theory will be the twin of a standard theory (i.e., have the same kink solution $\phi_k$ with the same energy density)  if both ${\cal E}$ and ${\cal P} \equiv 0$ agree when evaluated for the kink solution. A necessary and sufficient condition for the K field Lagrangian is 
\cite{trodden}
\bea \label{cond1a}
{\cal L}\vert_{\phi_k} &=& -2U \\
 {\cal L}_{,X} \vert_{\phi_k} &=& 1 , \label{cond2a} 
\eea
as may be checked easily.
Now the important point is that the first order form $\phi'^2 = -2X =2U$ of the static kink equation may be interpreted as an algebraic equation involving the variables $X$ and $\phi$ on which the K field Lagrangian depends. As a consequence, the evaluation condition $\vert_{\phi_k}$ may be replaced by the purely algebraic condition $\vert_{X=-U}$, leading to the so-called algebraic twin conditions \cite{twins-1}
\bea \label{cond1b}
{\cal L}\vert_{X=-U} \equiv {\cal L}\vert &=& -2U \\
 {\cal L}_{,X} \vert_{X=-U} \equiv {\cal L}_{,X} \vert &=& 1 \label{cond2b}
\eea 
(here and below the evaluation of an expression at $X\equiv -(1/2)\phi'^2 =-U$ (and its prolongations, when required) will always be denoted by the vertical line $\vert$, and will be called on-shell condition or on-shell evaluation frequently).

\section{The algebraic conditions}
\subsection{The fluctuation equation}
We start from the  Euler--Lagrange equation (\ref{eul-lag})
and insert the decomposition
\be 
\phi (t,x) = \phi_k (x) + \eta (t,x)
\ee
where $\phi_k$ is the kink solution and $\eta$ is the fluctuation  field. In first order in $\eta$ we find
\be
\pa_\mu \left( {\cal L}_{,X} \pa^\mu \eta + {\cal L}_{,XX} \pa_\nu \phi_k \pa^\nu \eta \pa^\mu \phi_k + {\cal L}_{,X\phi} \eta \pa^\mu \phi_k \right) - {\cal L}_{,\phi\phi} \eta - {\cal L}_{,X\phi} \pa_\mu \phi_k \pa^\mu \eta =0 .
\ee
Now we use the fact that $\phi_k$ only depends on $x$, and the ansatz for the fluctuation field
\be
\eta (t,x)= \cos (\omega t) \eta (x)
\ee
and get
\be
\left( - {\cal L}_{,X} \eta ' + {\cal L}_{,XX} (\phi_k ')^2 \eta ' - {\cal L}_{,X\phi} \phi_k ' \eta \right) ' -
{\cal L}_{,\phi \phi} \eta + {\cal L}_{,X\phi}\phi_k ' \eta ' - \omega^2 {\cal L}_{,X} \eta =0
\ee
 or, more explicitly
\bea
&& - \left( {\cal L}_{,X} + 2X{\cal L}_{,XX} \right) \eta '' - \left( {\cal L}_{,X\phi} +2X{\cal L}_{,XX\phi} - \phi_k '' (3{\cal L}_{,XX} +2X{\cal L}_{,XXX} ) \right) \phi_k ' \eta ' \nonumber \\
&=& \left( \omega^2 {\cal L}_{,X} + {\cal L}_{,\phi\phi} - 2 X {\cal L}_{,X\phi\phi} + \phi_k '' ({\cal L}_{,X\phi} +2X {\cal L}_{,XX\phi} ) \right) \eta .
\eea
This expression should now be evaluated for the defect solution $\phi_k$, i.e., implementing the on-shell condition $X\vert = -U$ and its first prolongation (that is, the original second order static field equation) $\phi ''\vert \equiv \phi_k '' = U_{,\phi}$. Inserting these on-shell expressions above produces an expression containing $U$ and its derivative, whereas the variables of ${\cal L}$  and its derivatives are $X\; (=-U)$ and $\phi$. The problem is that for a general potential $U$ the algebraic relation between $\phi$ and $U$ is undetermined, so we would have to treat each potential separately, losing thereby some of the generality of the algebraic method. The obvious alternative is to assume that the Lagrangian depends on     
$\phi$ only via the potential $U$, that is, ${\cal L} = {\cal L}(X,U)$. With 
\be
{\cal L}_{,\phi} = {\cal L}_{,U}U_{,\phi} \; , \quad {\cal L}_{,\phi\phi} = {\cal L}_{,UU}U_{,\phi}^2 + {\cal L}_{,U}U_{,\phi\phi}
\ee
we may rewrite the fluctuation equation like
\bea
&& - \left( {\cal L}_{,X} + 2X{\cal L}_{,XX} \right) \eta '' - \left( ({\cal L}_{,XU} +2X{\cal L}_{,XXU}) U_{,\phi} - \phi_k '' (3{\cal L}_{,XX} +2X{\cal L}_{,XXX} ) \right) \phi_k ' \eta ' = \nonumber \\
&& \left( \omega^2 {\cal L}_{,X} + {\cal L}_{,UU}U_{,\phi}^2  + {\cal L}_{,U} U_{,\phi\phi}- 2 X {\cal L}_{,XUU}U_{,\phi}^2 - 
2 X{\cal L}_{,XU}U_{,\phi\phi}  + \phi_k '' ({\cal L}_{,XU} +2X {\cal L}_{,XXU} ) U_{,\phi} \right) \eta \nonumber \\ &&
\eea
or, after implementing the on-shell conditions
\be 
X\vert = -U \; ,\quad \phi ''\vert = \phi_k '' = U_{,\phi} ,
\ee
like
\bea
&& - \left( {\cal L}_{,X} + 2X{\cal L}_{,XX} \right) \vert \, \eta '' - \left[ ({\cal L}_{,XU}  - 3{\cal L}_{,XX}  +2U ( {\cal L}_{,XXX}  - {\cal L}_{,XXU})  \right] 
\vert \, U_{,\phi} \phi_k ' \eta ' = \nonumber \\
&& \left[ \omega^2 {\cal L}_{,X} + U_{,\phi}^2 ({\cal L}_{,UU} +{\cal L}_{,XU}  + 2 U ( {\cal L}_{,XUU} - 
 {\cal L}_{,XXU} )) + U_{,\phi\phi} ( {\cal L}_{,U} + 2U {\cal L}_{,XU} )   \right] \vert \, \eta .
\eea
This expression should now be compared with the fluctuation equation of the standard case,
\be
-\eta '' = (\omega^2 - U_{,\phi\phi}\vert )\eta .
\ee
Comparing the standard and generalized fluctuation equations for a twin defect solution, and taking into account the twin condition ${\cal L}_{,X}\vert =1$, we find that a sufficient condition for the equality of the two fluctuation equations is provided by the following on-shell conditions
\be \label{fluct-cond1}
{\cal L}_{,XX}\vert =0
\ee
\be \label{fluct-cond2}
[{\cal L}_{,XU} +2U( {\cal L}_{,XXX} - {\cal L}_{,XXU}) ] \vert =0
\ee
\be \label{fluct-cond3}
[{\cal L}_{,UU} + {\cal L}_{,XU} + 2U({\cal L}_{,XUU} - {\cal L}_{,XXU}) ] \vert =0
\ee
and
\be \label{fluct-cond4}
({\cal L}_{,U} +2U{\cal L}_{,XU})\vert =-1.
\ee
These conditions are, again, purely algebraic conditions which the Lagrangian has to obey. If a Lagrangian obeys these conditions and the two twin conditions (\ref{cond1b}), (\ref{cond2b}), then it not only shares the same twin defect with the standard Lagrangian, but also the spectra of linear fluctuations about the defects coincide. 

\subsection{Examples}
It is easy to understand that there must exist infinitely many Lagrangians for each $U$ which obey these conditions. Indeed, if the Lagrangian ${\cal L}(X,U)$ is interpreted as a function of two independent variables $X$ and $U$, then the six twin and linear fluctuation conditions are just conditions which the first few Taylor coefficients of ${\cal L}$ must obey "on the diagonal", i.e., for $X=-U$. In a next step, let us construct, as a first example, a class of infinitely many Lagrangians which obey these conditions. These Lagrangians were, in fact, already introduced in \cite{twins-1} as examples of twins of the standard Lagrangian without noticing that they also give rise to coinciding fluctuation spectra. The class of Lagrangians is given by
\be \label{lag-ex}
{\cal L}^{\rm ex1} = \sum_{i=3,5,\ldots}^{2N+1} f_i (U) (X+U)^i + X - U \; , \quad f_i (U) \ge 0
\ee
where the $f_i$ are arbitrary nonnegative functions of their argument. The restriction to odd $i$ implies that the above Lagrangian obeys the null energy condition (NEC) and, therefore, defines a healthy (stable) field theory. We remark that this restriction may be relaxed without violating the NEC provided that the $f_i$ for even $i$ obey certain inequalities, but here we restrict to odd $i$ for reasons of simplicity. It is easy to check that the above Lagrangian obeys
\be
{\cal L}^{\rm ex1}_{} \vert =-2U ; , \quad  {\cal L}^{\rm ex1}_{,X} \vert = 1
\ee
i.e., the twin conditions, as well as
\be 
{\cal L}^{\rm ex1}_{,XX} \vert = {\cal L}^{\rm ex1}_{,XU} \vert = {\cal L}^{\rm ex1}_{,UU} \vert = 0 \;  ,\quad {\cal L}^{\rm ex1}_{,U} \vert = -1
\ee
and
\be
{\cal L}^{\rm ex1}_{,XXX} \vert = {\cal L}^{\rm ex1}_{,XXU} \vert = {\cal L}^{\rm ex1}_{,XUU} \vert = 6f_3.
\ee
Further, these conditions obviously imply the "fluctuation conditions" (\ref{fluct-cond1}) - (\ref{fluct-cond4}), therefore the class of Lagrangians (\ref{lag-ex}) not only are twins of the standard Lagrangian ${\cal L}_s = X-U$ (i.e. they share the same kink solution with the same energy density), but also the linear fluctuation spectra about the kink solutions coincide.

We remark that it is obvious from the above derivation that the restriction to $f_i = f_i (U)$ in the above class of examples is not necessary, and we may in fact allow for functions $f_i = f_i (\phi) \ge 0$ without changing our results. 

Another class of examples is provided by the power series expansion
\be
{\cal L}^{\rm ex2} = \sum_{i=0,j=0}^{M,N}a_{ij}X^i(X+U)^j  - 2U
\ee
where the twin and fluctuation conditions lead to 
\be
a_{0j}=0 \; \forall \; j \; , \quad a_{10}=1 \; , \quad a_{1j} =0 \; , j=1\ldots N \; , \quad a_{2j}=0 \; \forall \; j .
\ee
It is again possible to satisfy the NEC by imposing the corresponding conditions (inequalities) on the nonzero coefficients $a_{ij}$.

For a more systematic search for examples it is useful to perform the following transformation of variables,
\be
Y=X+U, \quad Z=U \qquad \Rightarrow \qquad \pa_X = \pa_Y ,\quad \pa_U = \pa_Y + \pa_Z
\ee
where the evaluation condition now means evaluation at $Y=0$, i.e., $\vert \equiv \vert_{Y=0}$. Shifting, in addition, the lagrangian by $2U$,
\be
\tilde{\cal L} = {\cal L} +2U
\ee
the two twin conditions and the first fluctuation condition read
\be \label{twin1t}
\tilde{\cal L}\vert =0
\ee
\be \label{twin2t}
\tilde{\cal L}_{,Y} \vert =1
\ee
and 
\be \label{fluct1t}
\tilde{\cal L}_{,YY} \vert =0
\ee  
and, taking these into account, the remaining fluctuation conditions become
\be \label{fluct2t}
\left( \tilde{\cal L}_{,Z} +2Z \tilde{\cal L}_{,YZ} \right) \vert =0
\ee
\be \label{fluct3t}
\left( \tilde{\cal L}_{,YZ} -2Z\tilde{\cal L}_{,YYZ} \right) \vert =0
\ee
and
\be \label{fluct4t}
\left[ 2\tilde{\cal L}_{,YZ} + \tilde{\cal L}_{,ZZ} + 2Z(\tilde{\cal L}_{,YYZ} + \tilde{\cal L}_{,YZZ}) \right] \vert =0.
\ee
As an application, let us study the Dirac--Born--Infeld (DBI) type theory which was first introduced in \cite{trodden} as an example for a K field twin,
\bea
\tilde{\cal L}^{\rm DBI} &=& -\sqrt{1+2U}\sqrt{1-2X} + \sum_if_i (U) (X+U)^i \nonumber \\
&=& -\sqrt{1+2Z}\sqrt{ 1-2Y+2Z} + \sum_i f_i(Z) Y^i
\eea
where the task consists in determining the coefficient functions $f_i (Z)=f_i(U)$ such that all the twin and fluctuation conditions are satisfied.
After some calculation one finds that the two twin conditions (\ref{twin1t}), (\ref{twin2t}) and the first fluctuation condition (\ref{fluct1t}) lead to
\be
f_0 = 1+2Z \; , \quad f_1 =0 \; , \quad f_2 = \frac{1}{2}\frac{1}{1+2Z}
\ee
whereas the remaining fluctuation conditions are satisfied identically precisely for the above solutions for $f_0$, $f_1$ and $f_2$. We conclude that the DBI type Lagrangian 
\be
{\cal L}^{\rm DBI} = -\sqrt{1+2U}\sqrt{1-2X} +1+\frac{1}{2}\frac{1}{1+2U}(X+U)^2
\ee
is a twin of the standard Lagrangian $X-U$ with coinciding linear fluctuation spectra about the common (twin) defect solution. 
The above DBI type Lagrangian as it stands does not obey the NEC, but we are allowed to add, e.g., a cubic term $f_3(X+U)^3$ without altering the twin or fluctuation conditions. It may be checked easily that, e.g., for functions $f_3(U)$ obeying the inequality $f_3 \ge [1/(3(1+2U)^2)]$, the resulting Lagrangian does obey the NEC. 

Obviously, our algebraic method may be used without difficulty to produce more examples of K field twins with coinciding linear fluctuation spectra.

\subsection{The examples of Bazeia and Menezes}
In their recent paper \cite{twins-bazeia2}, Bazeia and Menezes introduced a class of Lagrangians given by the following ansatz,
\be \label{lag-BM}
{\cal L}^{\rm BM} = - UF(Y) \; ,\quad Y\equiv -\frac{X}{U}
\ee
where $F$ is an arbitrary function of its argument. This ansatz may be justified by the observation that both the twin conditions (\ref{cond1b}), (\ref{cond2b}) and the fluctuation conditions (\ref{fluct-cond1}) - (\ref{fluct-cond4}) are compatible with a Lagrangian which is a homogeneous function of degree one in its two variables $X$ and $U$.  The Lagrangian in (\ref{lag-BM}) obviously is such a homogeneous function of degree one.
For the partial derivatives w.r.t $X$ and $U$ we get
\be
{\cal L}^{\rm BM}_{X} = F' \; , \quad {\cal L}^{\rm BM}_{XX} = -\frac{F''}{U} \; ,\quad {\cal L}^{\rm BM}_{XXX} = \frac{F'''}{U^2}
\ee
\be
{\cal L}^{\rm BM}_{U} = -F-\frac{X}{U}F' \; ,\quad {\cal L}^{\rm BM}_{UU} = -\frac{X^2}{U^3} F''
\ee
and
\be
{\cal L}^{\rm BM}_{XU} = \frac{X}{U^2} F'' \; ,\quad {\cal L}^{\rm BM}_{XUU} = -2\frac{X}{U^3}F'' + \frac{X^2}{U^4}F''' \; ,\quad
{\cal L}^{\rm BM}_{XXU} = \frac{F''}{U^2} -\frac{X}{U^3} F''' .
\ee
These expressions should now be evaluated on-shell, i.e., for $X=-U$, and inserted into the twin and fluctuation conditions. We shall find that the homogeneity of the ansatz (\ref{lag-BM})  not only is compatible with these conditions, but also leads to a considerable simplification for the fluctuation conditions. First of all, for the twin conditions we find 
\be
{\cal L}^{\rm BM} \vert = -UF(1) = -2U \quad \Rightarrow \quad F(1)=2
\ee  
and 
\be
{\cal L}^{\rm BM}_{,X} \vert = F'(1) = 1
\ee  
where the on-shell condition $X=-U$ implies that the function $F(Y)$ and its derivatives are evaluated at $Y=1$. For the fluctuation conditions we find that condition (\ref{fluct-cond3}) is satisfied identically without providing a further restriction, whereas the remaining conditions lead to
\be
{\cal L}^{\rm BM}_{XX} \vert = -\frac{F''(1)}{U}  = 0 
\ee
\be 
[{\cal L}_{,XU} +2U( {\cal L}_{,XXX} - {\cal L}_{,XXU}) ] \vert =- \frac{2}{U}F''(1) = 0 
\ee
and 
\be
({\cal L}_{,U} +2U{\cal L}_{,XU})\vert =-F(1) + F' (1) - 2 F''(1) = -1 -2 F''(1) = -1
\ee
where we used the two twin conditions in the last expression. In other words, for the ansatz of Bazeia and Menezes, all four fluctuation conditions just boil down to the simple condition 
\be
F'' (1)=0.
\ee 

Finally, Bazeia and Menezes gave the following explicit example (one-parameter family of Lagrangians)
\be
F(Y) = 1+Y+\frac{\alpha}{3} (1-Y)^3 \quad \Rightarrow \quad {\cal L}^{BM,\alpha} = X-U + \frac{\alpha}{3U^2}(X+U)^3
\ee
where $\alpha$ is a real, positive constant. This example
belongs, in fact, to the first class of examples discussed in the previous subsection. Concretely it is of the type
(\ref{lag-ex}) for the choice 
\be
f_3 (U) = \frac{\alpha}{3U^2} \; ,\quad f_i =0 \quad \mbox{for} \quad i>3.
\ee 

\section{Conclusions}
In this article we demonstrated that for every standard scalar field theory ${\cal L}_s =X - U(\phi)$ which supports a topological defect (a kink), there exist infinitely many generalized (or K) field theories ${\cal L}( X, \phi)$ ("twins" of the standard field theory) which support the same kink with the same energy density and with the same spectrum of linear fluctuations about the kink.
Further, we gave a
simple and explicit algebraic method to construct these twins of the standard scalar field theory with identical linear fluctuation spectra. As stated, some first examples of such twinlike models with coinciding kink solutions, energy densities and linear fluctuation spectra have been given already in \cite{twins-bazeia2}. K field twin defects with coinciding linear fluctuation spectra are almost completely indistinguishable from their standard counterparts and, as a consequence, the K field theories giving rise to them have to be considered on a par with the standard field theories in all situations where K field theories cannot be excluded on theoretical grounds. In particular, in the context of effective field theories, where higher kinetic terms are induced naturally, the topological defects formed in K field theories should be taken as seriously as their standard field theory   
twins, because they give rise to almost exactly the same physics. 
In this context, an observation of special interest is related to the fact that the coinciding linear fluctuation spectra imply that a semiclassical quantization about the topological defect provides the same results for the standard defect and its K field twins. This not only facilitates specific physical properties of the K field defect, but, more generally, provides us with a first partial result on the quantization of K field theories, which, in general,  is a still unsolved and probably quite difficult problem.

Finally, let us briefly comment on possible generalizations and future work. A first issue is the inclusion of fermions and the supersymmetric extension of K field twins. Supersymmetric (SUSY) extensions of scalar K field theories have been found recently \cite{susy-bazeia}, \cite{susy2}, \cite{susy-bS}, and some examples of SUSY K field twins of standard SUSY theories have been given already in \cite{twins-1}. Here, one interesting question obviously is what the coinciding fluctuation spectra in the twin kinks imply for the SUSY fermions. Another interesting generalization concerns the issue of twins of topological defects in higher dimensions, like, e.g., vortices, monopoles, or skyrmions, possibly  after a symmetry reduction (e.g. to spherical symmetry) of the Lagrangian or Euler--Lagrange equations. The case of vortices in generalized abelian Higgs models has been investigated in the very recent paper \cite{twins-bazeia3}, where the authors do find twins of standard vortices. Certainly these issues are worth further investigation.
 
\section*{Acknowledgement}
The authors acknowledge financial support from the Ministry of Science and Investigation, Spain (grant FPA2008-01177), 
the Xunta de Galicia (grant INCITE09.296.035PR and
Conselleria de Educacion), the
Spanish Consolider-Ingenio 2010 Programme CPAN (CSD2007-00042), and FEDER. Further, the authors thank J. Sanchez-Guillen for helpful discussions.


\begin{thebibliography}{100}
\bb{man-sut-book}
N. Manton, P. Sutcliffe, "Topological Solitons", Cambridge University Press, Cambridge, 2007.
\bibitem{Vil}
A. Vilenkin, E.P.S. Shellard, {\em Cosmic strings and other topological defects} Cambridge University Press, 1994.
\bibitem{Hind}
M. Hindmarsh, T.W.B. Kibble, Rep. Prog. Phys. {\bf 58}, 477 (1995).
\bibitem{Battye}
R.A. Battye, J. Weller, Phys. Rev. D{\bf 61}, 043501 (2000). 
\bb{Go-Jac}
J. Goldstone, R. Jackiw, Phys. Rev. D{\bf 11}, 1486 (1975).
\bb{raja-book}
R. Rajaraman, "Solitons and Instantons", Elsevier Science, Amsterdam, 1982.
\bb{k-infl}
C. Armendariz-Picon, T. Damour,  V. Mukhanov, Phys. Lett. B{\bf458}, 209 (1999) [hep-th/9904075].
\bb{k-ess}C. Armendariz-Picon, V. Mukhanov, P.J. Steinhardt, Phys. Rev. Lett. {\bf85}, 4438 (2000) [astro-ph/0004134];
C. Armendariz-Picon, V. Mukhanov, P.J. Steinhardt, Phys. Rev. D{\bf63}, 103510 (2001) [astro-ph/0006373].
\bb{comp-brane} C. Adam, N. Grandi, J. Sanchez-Guillen, A. Wereszczynski, J. Phys. A{\bf41}, 212004 (2008) [arXiv:0711.3550];
C. Adam, N. Grandi, P. Klimas, J. Sanchez-Guillen, A. Wereszczynski, J. Phys. A{\bf41}, 375401 (2008) [arXiv:0805.3278].
\bb{olech}M. Olechowski, Phys. Rev. D{\bf78}, 084036 (2008) [arXiv:0801.1605].
\bb{brane-bazeia}
D. Bazeia, A. R. Gomes, L. Losano, R. Menezes, Phys. Lett. B{\bf 671}, 402 (2009) [arXiv:0808.1815].
\bb{babichev1} E. Babichev, Phys. Rev. D{\bf 74}, 085004 (2006) [hep-th/0608071].
\bb{nicole-mod}
C. Adam, J. Sanchez-Guillen, R.A. Vazquez, A. Wereszczynski,
 J. Math. Phys. {\bf 47}, 052302 (2006) [hep-th/0602152].
\bb{werle}
J. Werle,
Phys. Lett. B{\bf 71}, 367 (1977).
\bb{rosenau1}P. Rosenau, J. M. Hyman, Phys. Rev. Lett. {\bf70}, 564 (1993); P. Rosenau, Phys. Rev. Lett. {\bf73}, 1737 (1994).
\bb{sodano1}
F. Cooper, H. Shepard, P. Sodano, 
Phys. Rev. E{\bf 48}, 4027 (1993);
B. Mihaila, A. Cardenas, F. Cooper, A. Saxena, Phys. Rev. E{\bf 82}, 066702 (2010).
\bb{arodz}
H. Arodz, Acta Phys. Polon. B{\bf 33}, 1241 (2002); H. Arodz, P. Klimas, T.
Tyranowski, Acta Phys. Polon. B{\bf 36}, 3861 (2005).
\bb{comp}
C. Adam, J. Sanchez-Guillen, A. Wereszczynski, J. Phys. A{\bf40}, 13625 (2007) [arXiv:0705.3554]; Erratum-ibid. A{\bf42}, 089801 (2009).
\bb{comp-vert}
C. Adam, P. Klimas, J. Sanchez-Guillen, A. Wereszczynski, J. Phys. A{\bf 42}, 135401 (2009).
\bb{bazeia4}
D. Bazeia, E. da Hora, R. Menezes,  H.P. de Oliveira, C. dos Santos, 
Phys. Rev. D{\bf 81}, 125016 (2010) 
[arXiv:1004.3710].
\bb{dosSantos1}
C. dos Santos, Phys. Rev. D{\bf 82}, 125009 (2010).
\bb{comp-bS}
T. Gisiger, M.B. Paranjape, Phys. Rev. D{\bf 55}, 7731 (1997); 
C. Adam, P. Klimas, J. Sanchez-Guillen, A. Wereszczynski, Phys. Rev. D{\bf 80}, 105013 (2009); 
C. Adam, T. Romanczukiewicz,  J. Sanchez-Guillen, A. Wereszczynski, 
Phys. Rev. D{\bf 81}, 085007 (2010);
J.M. Speight, 
J. Phys. A{\bf 43}, 405201 (2010). 
\bb{fring}
P. E. G. Assis, A. Fring, Pramana  J. Phys. {\bf 74} (2010) 857.
\bb{bab-muk-1}
E. Babichev, V. Mukhanov, A. Vikman,
JHEP {\bf 0802}, 101 (2008) 
[arXiv:0708.0561]
\bb{bergli1}
E. Goulart, S.E. Perez Bergliaffa, 
arXiv:1108.3237 
\bb{trodden}
M. Andrews, M. Lewandowski, M. Trodden, D. Wesley, Phys. Rev. D{\bf 82}, 105006 (2010).
\bb{twins-bazeia1}
D. Bazeia,  J.D. Dantas,  A.R. Gomes, L. Losano, R. Menezes, 
Phys. Rev. D{\bf 84}, 045010 (2011); 
arXiv:1105.5111 [hep-th].
\bb{twins-1}
C. Adam, J.M. Queiruga, Phys. Rev. D{\bf 84}, 105028 (2011)
[arXiv:1109.4159].
\bb{twins-bazeia2}
D. Bazeia, R. Menezes, arXiv:1111.1318 [hep-th].
\bb{bazeia3}
%% Generalized Global Defect Solutions.
D. Bazeia, L. Losano, R. Menezes, J.C.R.E. Oliveira
Eur. Phys. J. C{\bf 51}, 953 (2007),
[hep-th/0702052];
D. Bazeia, L. Losano, R. Menezes, Phys. Lett. B{\bf 668}, 246 (2008), [arXiv:0807.0213].
\bb{susy-bazeia} 
D. Bazeia, R. Menezes,  A.Yu. Petrov, 
Phys. Lett. B{\bf 683}, 335 (2010) 
[arXiv:0910.2827]. 
\bb{susy2}
C. Adam, M. Queiruga, J. Sanchez Guillen, A. Wereszczynski, Phys. Rev. D{\bf 84}, 065032 (2011)
[arXiv:1107.4370].
\bibitem{susy-bS}
C. Adam, M. Queiruga, J. Sanchez Guillen, A. Wereszczynski,
Phys. Rev. D{\bf 84}, 025008 (2011)
[arXiv:1105.1168].
\bb{twins-bazeia3}
D. Bazeia, E. da Hora, R. Menezes, arXiv:1111.6542.

\end{thebibliography}
\end{document}